\documentclass[12pt]{iopart}

\usepackage{multirow}
\usepackage{graphicx}% Include figure files
\usepackage{dcolumn}% Align table columns on decimal point
\usepackage{bm}% bold math

\begin{document}

\title{Three-way noiseless signal splitting in a parametric amplifier with quantum correlation}

\author{Nannan Liu$^{1}$, Jiamin Li$^{1}$, Xiaoying Li$^{1 *}$, Z. Y. Ou$^{1,2\dagger}$}

\address{$^{1}$ College of Precision Instrument and
Opto-electronics Engineering, Tianjin University, \\Key Laboratory of Optoelectronics Information Technology, Ministry of Education, Tianjin, 300072,
P. R. China\\$^{2}$ Department of Physics, Indiana University-Purdue University Indianapolis, Indianapolis, IN 46202, USA}
\ead{$^{*}$xiaoyingli@tju.edu.cn $^{\dagger}$zou@iupui.edu}

\begin{abstract}
We demonstrate that a phase-insensitive parametric amplifier, coupled to a quantum correlated source, can be used as a  quantum information tap for noiseless three-way signal splitting. We find that the output signals are amplified noiselessly in two of the three output ports while the other can more or less keep its original input size without adding noise. This scheme is able to cascade and scales up for efficient information distribution in an optical network. Furthermore, we find this scheme satisfies the criteria for a non-ideal quantum non-demolition (QND) measurement and thus can serve as a QND measurement device. With two readouts correlated to the input,  we find this scheme also satisfies the criterion for sequential QND measurement.
\end{abstract}

%\submitto{\NJP}
%\maketitle

\section{Introduction}
\noindent Distribution of information without adding noise is the goal of an information network. It is not a problem to accomplish this in a classical system since we can make multiple copies nearly noiselessly. For a quantum network, however, such a task met a tremendous challenge: an arbitrary quantum state cannot be copied with a universal cloning machine due to quantum no cloning theorem \cite{noclone}.  On the other hand, this theorem does not prevent us from cloning a specific state, e.g., the eigen-states of the system. This is the underlying principle for noiseless quantum amplification \cite{qu-psa}. Even with  the noiseless amplification, we still need to split the information for distribution.
But this task cannot be achieved with simple linear beam splitters because of the quantum noise of vacuum that is coupled in via the unused port \cite{sha}. It was suggested \cite{sha} and implemented \cite{bru97-1} that squeezed states are injected in the unused port for quantum optical information tapping. But the scheme with a beam splitter always reduces the  size  of the signal and the readouts. Further noiseless amplification is thus required for  information distribution, leading to a complicated configuration that is hard to scale up for integration.

It was demonstrated recently that a parametric amplifier coupled with correlated quantum source functions as a quantum information tap that can split a quantum signal better than a classical beam splitter and in the meantime amplify the signal \cite{guo}. Unfortunately, this scheme cannot be cascaded and is not suitable for information distribution.
In this paper, we analyze a scheme that combines the beam splitter scheme and the scheme for noiseless amplification with quantum correlation. We find that it can be used for three-way noiseless signal splitting with two of the outputs amplified and the third almost at input level and thus it is possible for cascade and scale-up.

It should be mentioned that some other schemes of quantum information tapping are based on quantum non-demolition measurement (QND) \cite{bru97-1,poi93,per94,ben95,poi96,bru97-0,bru97,wang99}. On the other hand, the ability to perform quantum information tapping does not guarantee the scheme to be a QND measurement\cite{bru97}. As discussed by Holland et al \cite{hol}, a set of three criteria must be satisfied for a non-ideal QND measurement. Our proposed scheme will be tested under these criteria. Furthermore, for a sequential QND measurement scheme\cite{ben95,bru97}, another criterion \cite{bru97} is needed for non-ideal state projection in order for it to become a QND measurement.

The paper is organized as follows: In section 2, we discuss quantum optical tapping with beam splitters and correlated quantum sources. In section 3, we analyze a recent noiseless amplification scheme with non-degenerate optical parametric amplifier for quantum information tapping. In section 4, we combine the two schemes in sections 2 and 3 to achieve three-way noiseless signal splitting. In section 5, we discuss the conditions for the three-way splitting scheme to satisfy the QND measurement criteria. We conclude with a summary and discussion.

\section{Quantum information tapping with a beam splitter coupled to a squeezed vacuum or a quantum correlated source}

First, let us consider a scheme of quantum information tapping with a beam splitter. With a squeezed state input at the unused port, as shown in Fig. 1(a), this is the scheme proposed by Shapiro \cite{sha}. For an input coherent state $|\alpha\rangle$ with a non-zero quadrature-phase amplitude $\langle \hat X_{in} \rangle = 2\alpha$ $(\alpha=real,~ \hat X_{in} = \hat a_{in}+\hat a_{in}^{\dag})$, we have the signal-to-noise ratio (SNR) of the input as
\begin{eqnarray}
R_{in} = \langle \hat X_{in} \rangle^2/\langle \Delta^2\hat X_{in} \rangle = 4\alpha^2,
\label{SNRi}
\end{eqnarray}
where the noise for the coherent state $\langle \Delta^2\hat X_{in} \rangle =1 $ is simply the vacuum noise. For the two outputs of the beam splitter with a transmissivity $T$, it is straightforward to show \cite{sha} that the SNRs are
\begin{eqnarray}
R_1= {T(2\alpha)^2\over T+(1-T)S}, ~~ R_2 = {(1-T)(2\alpha)^2\over 1-T +TS},
\label{SNRo}
\end{eqnarray}
respectively. Here the numerators are the signal whereas the denominators are the noise. $S$ is the degree of squeezing for the squeezed state ($S=1$ for vacuum and $S=0$ for perfect squeezing). The information transfer can be characterized by the transfer coefficients for the two output ports:
$
{\cal T}_{1,2} = R_{1,2}/R_{in}
$
which is simply the inverse of the noise figure of the system.
Therefore the total transfer coefficient for the signal splitting with a beam splitter is
\begin{eqnarray}
{\cal T}_{BS} = R_1/R_{in}+ R_2/R_{in} = {T\over T+(1-T)S} + {1-T\over TS+ 1-T},
\label{Tbs}
\end{eqnarray}
which reaches a maximum value of $2/(1+S)$ when $T=1/2$. Notice that when $S=1$ for vacuum input at the unused port (no squeezing), we have ${\cal T}_{BS}=1$ indicating no information tapping, i.e., the split signals have reduced transfer coefficients due to the addition of vacuum noise. But when squeezing is perfect with $S=0$, we have ${\cal T}_{BS}=2$ for complete information tapping with the signal splitting into two without degradation of the transfer coefficients. Another feature in this scheme is that the signal sizes in the outputs are reduced by half in the optimum case of $T=1/2$.

\begin{figure}[h]
\includegraphics[width=5in]{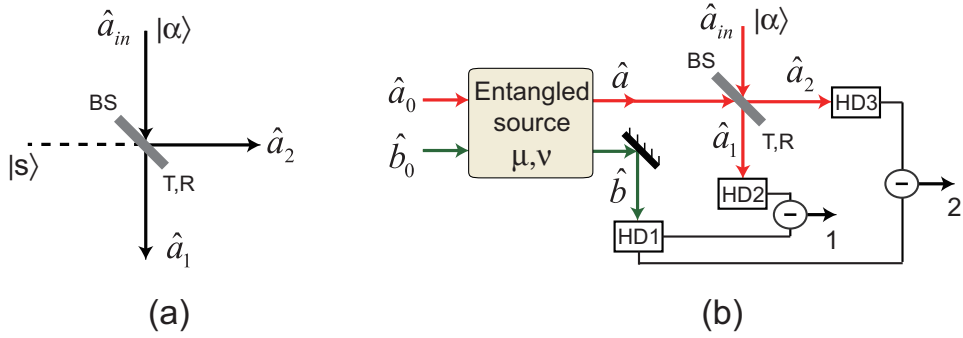}
\caption{Quantum information taps using a beam splitter with the unused port injected by (a) a squeezed vacuum or (b) a correlated quantum entangled source. BS: beam splitter; HD: homodyne detection.}
\end{figure}

Quantum correlated sources also lead to quantum noise reduction and can be arranged for quantum information tapping, as shown in Fig. 1(b). Assume that $\hat a$ and $\hat b$ are the two quantum entangled fields from a non-degenerate parametric amplifier that satisfies the Bogoliubov transformation from vacuum modes $\hat a_0,~\hat b_0$:
\begin{eqnarray}
\hat a &=& \mu \hat a_0 + \nu\hat b_0^{\dag},\cr
\hat b &=& \mu \hat b_0 + \nu\hat a^{\dag}_0,
\label{b12}
\end{eqnarray}
where $\mu,~\nu$ are the amplitude gain coefficients of the non-degenerate parametric amplifier for the entangled source ($\mu^2-\nu^2=1$). The quantum noise of the fields are correlated and can be reduced by subtraction. As before in Fig. 1(a), the input signal field $\hat a_{in}$ in Fig. 1(b) is in a coherent state $|\alpha \rangle$  with a non-zero quadrature-phase amplitude  $\langle \hat X_{in} \rangle = 2\alpha~ (\alpha=real)$. The outputs from the beam splitter with a transmissivity $T$ are
\begin{eqnarray}
\hat a_1 &=& \hat a_{in} \sqrt{T} +\hat a\sqrt{1-T},\cr
\hat a_2 &=& \hat a \sqrt{T} -\hat a_{in}\sqrt{1-T}.
\label{a12}
\end{eqnarray}
Because the noise of $\hat a$ is correlated with $\hat b$, we need to subtract the current from homodyne measurement of $\hat b$. So, the outputs (1 and 2) are $\hat X_1 = \hat X_{a_1} - \lambda_1\hat X_{b}$ and $\hat X_2 = \hat X_{a_2} - \lambda_2\hat X_{b}$, respectively. Here $\lambda_{1,2}$ are the electronic gains for optimum noise reduction. The output signals are simply $\langle \hat X_1\rangle = \sqrt{T}(2\alpha),~ \langle \hat X_2\rangle = \sqrt{1-T}(2\alpha)$. After subtracting the noise from $\hat b$-field for quantum noise reduction, the noise of the two outputs are
\begin{eqnarray}
\langle \Delta^2\hat X_{1}\rangle &=& \big\langle (\sqrt{T}\Delta \hat X_{a_{in}} +\sqrt{1-T} \Delta \hat X_{a}- \lambda_1 \Delta\hat X_{b})^2 \big\rangle,\cr
\langle \Delta^2\hat X_{2}\rangle &=& \big\langle ( - \sqrt{1-T}\Delta \hat X_{a_{in}} +\sqrt{T} \Delta \hat X_{a}- \lambda_2 \Delta \hat X_{b})^2 \big\rangle.
\label{N12}
\end{eqnarray}
It is straightforward to find
\begin{eqnarray}
\langle \Delta^2\hat X_{1}\rangle &=& T + (1-T)S_e,\cr
\langle \Delta^2\hat X_{2}\rangle &=& 1-T + T S_e.
\label{N12-2}
\end{eqnarray}
with $\lambda_1=  2\mu\nu\sqrt{1-T}/(\mu^2+\nu^2),~ \lambda_2=  2\mu\nu\sqrt{T}/(\mu^2+\nu^2)$ for optimum noise reduction and $S_e\equiv 1/(\mu^2+\nu^2)$ for the degree of quantum noise reduction. We then have the signal-to-noise ratios for the two outputs:
\begin{eqnarray}
R_1= {T(2\alpha)^2\over T+(1-T)S_e}, ~~ R_2 = {(1-T)(2\alpha)^2\over T S_e+(1-T)}.
\label{SNRo-2}
\end{eqnarray}
Note that the expressions above are exactly the same as Eq. (\ref{SNRo}) for the scheme in Fig. 1(a) and thus lead to the same overall information transfer coefficient in Eq. (\ref{Tbs}).

\section{Non-degenerate parametric amplifier for noiseless amplification and quantum information tapping}

In the schemes involving a beam splitter, we found that the signal sizes are reduced by a factor of $T$ and $1-T$, which is undesirable if the schemes are used for cascade and scale up. To compensate the signal loss, we can resort to noiseless amplification by a degenerate parametric amplifier. But this adds some degree of complexity for the scheme.

Recently, noiseless amplification is achieved in a non-degenerate parametric amplifier (NPA) with the help of quantum correlated sources for  both the signal and the internal mode of the amplifier \cite{guo,kong13}. It was suggested \cite{ou93} and demonstrated \cite{guo} that the other idler output port of the amplifier should also contain the information about the input signal and can be used for quantum information tapping. We now analyze this scheme which is depicted in Fig. 2(a).

\begin{figure}[h]
\includegraphics[width=6in]{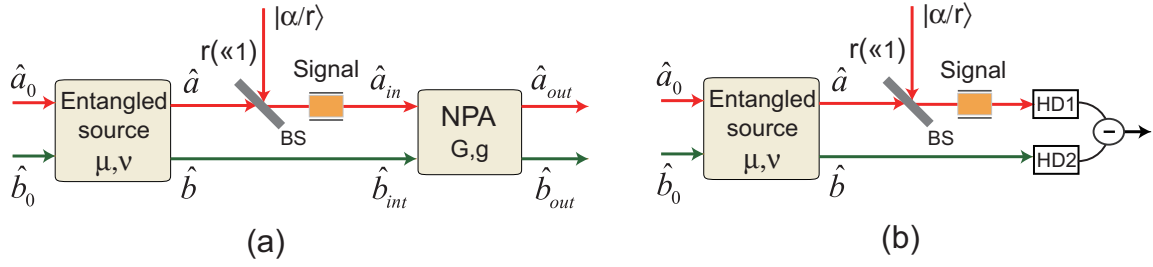}
\caption{(a) Quantum information tap using a non-degenerate parametric amplifier (NPA) with a correlated quantum entangled source input. (b) Dual beam detection of the input signal. BS: beam splitter; HD: homodyne detection.}
\end{figure}

A NPA can be described by
\begin{eqnarray}
\hat a_{out} = G \hat a_{in} + g\hat b_{int}^{\dag}
\label{aout}
\end{eqnarray}
where $G$ is the amplitude gain of the amplifier with $G^2-g^2=1$ and $\hat b_{int}$ is the internal mode of the amplifier. It has been shown in in Ref. \cite{ou93} that  when the input mode $\hat a_{in}$ and the internal mode $\hat b_{int}$ of the amplifier are quantum mechanically correlated (in a way similar to Eq. (\ref{b12})), the signal output port $\hat a_{out}$ of the NPA has the same SNR as the input SNR for dual beam detection if the amplifier's amplitude gain $G=\mu^2+\nu^2$ with $\mu,~\nu$ for the characterization of  the quantum correlation. For the idler output port, we have
\begin{eqnarray}
\hat b_{out} = G \hat b_{int}  + g\hat a_{in}^{\dag}.
\label{bout}
\end{eqnarray}
For the input signal, we couple a coherent state $|(\alpha/r)\rangle$ with one of the correlated sources (field $\hat a$) by a beam splitter with amplitude reflectivity $r \ll 1$ (Fig. 2(a)) so that the noise part is dominated by the correlated source while the signal part is a small part from the coherent state $\langle X_{in} \rangle = 2r(\alpha/r) = 2\alpha$ (The correlated source has a zero mean value). The output signal size for the idler mode is $\langle X_{b_{out}} \rangle = (2\alpha)g $ and the noise of the idler output is
\begin{eqnarray}
\langle \Delta^2 \hat X_{b_{out}}\rangle &=& \langle ( G \Delta \hat X_{b_{int}}  + g \Delta \hat X_{a_{in}})^2\rangle \cr
&=& (G^2+g^2)(\mu^2+\nu^2) - 4Gg\mu\nu.
\label{bN}
\end{eqnarray}
Note that when we calculate the above, we let $\hat a_{in}$ and $\hat b_{int}$ be correlated like $\hat a,~\hat b$ given in Eq. (\ref{b12}). The output signal-to-noise ratio is then
\begin{eqnarray}
R_b = {g^2(2\alpha)^2\over (G^2+g^2)(\mu^2+\nu^2) - 4Gg\mu\nu}.
\label{Rb}
\end{eqnarray}
From Ref. \cite{ou93}, we find the SNR for the signal output port as
\begin{eqnarray}
R_a = {G^2(2\alpha)^2\over (G^2+g^2)(\mu^2+\nu^2) - 4Gg\mu\nu}.
\label{Ra}
\end{eqnarray}
For the  dual beam detection (Fig. 2(b)), the  input SNR is simply $R_{in} = (2\alpha)^2(\mu^2+\nu^2)$. So, the overall transfer coefficient is
\begin{eqnarray}
{\cal T}= {R_a\over R_{in}} + {R_b\over R_{in}}= {(G^2+g^2)/(\mu^2+\nu^2)\over (G^2+g^2)(\mu^2+\nu^2) - 4Gg\mu\nu}.
\label{NF2}
\end{eqnarray}
This is optimized when $G=\mu^2+\nu^2$, which gives $R_a=R_{in}=(2\alpha)^2(\mu^2+\nu^2) = G(2\alpha)^2$ and
\begin{eqnarray}
R_b = {g^2(2\alpha)^2/G}.
\label{Rb}
\end{eqnarray}
Therefore, the overall transfer coefficient for the non-degenerate amplifier with correlated quantum source is
\begin{eqnarray}
{\cal T}_{NPAC}= 1+ g^2/G^2,
\label{NF3}
\end{eqnarray}
which is larger than 1 and approaches 2 for large gain $G\gg 1$ and can serve for quantum information tapping.

\section{Three-way quantum information splitting}

Although the scheme in the last section leads to amplified signal output, it is assumed that the input signal is already incorporated in the correlated source by some means (the modulator before field $\hat a_{in}$ in Fig. 2(a)). Thus it cannot be applied to an arbitrary input signal. In this section, we modify this scheme and combine it with the beam splitter scheme discussed in section 2 to form a three-way quantum information splitter.

\begin{figure}[h]
\includegraphics[width=6in]{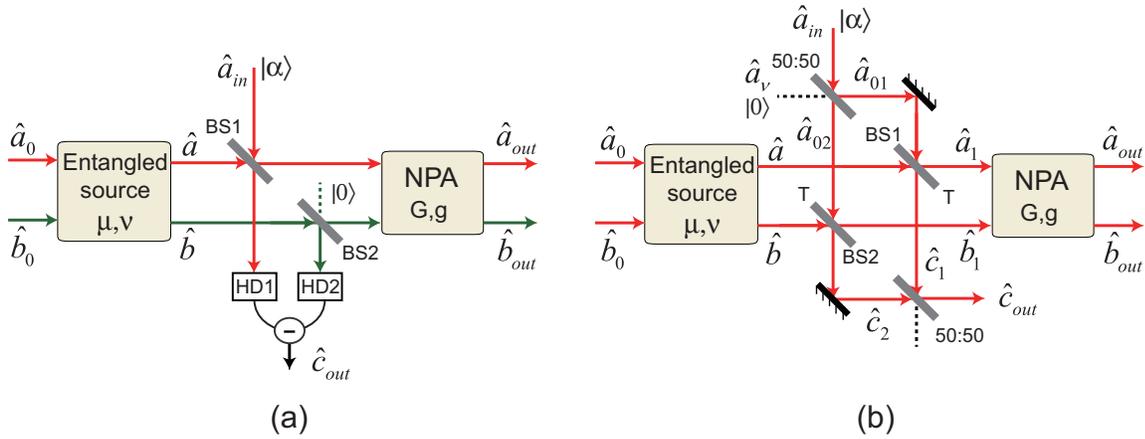}
\caption{Three-way quantum information splitting with a non-degenerate parametric amplifier (NPA) and a correlated quantum source by (a) one tapping beam splitter or (b) two tapping beam splitters. BS: beam splitter; HD: homodyne detection.}
\end{figure}

In order to work with an arbitrary input signal, we can encode the information in the coherent state before the beam splitter in Fig. 2. We also need to split part of the correlated field $\hat b$ to subtract the noise in the transmitted signal from the beam splitter and we obtain the simple scheme of one tapping beam splitter (BS1) in Fig. 3(a).
However, this scheme does not work because of the vacuum noise input at BS2. It can be shown that the total transfer coefficient for the three outputs is at most 3/2. To avoid vacuum noise, we use the scheme of two tapping beam splitters (BS1 and BS2) in Fig. 3(b) by splitting the input signal into two equal parts:
\begin{eqnarray}
\hat a_{01} = {\hat a_{in} + \hat a_{v}\over \sqrt{2}}, ~~~~\hat a_{02} = {\hat a_{in} - \hat a_{v}\over \sqrt{2}}.
\end{eqnarray}
Here $\hat a_v$ is the vacuum field introduced for input signal splitting. The effect of $\hat a_v$ is negligible for this scheme (see the discussion at the end). We then send the split signals to BS1 and BS2, respectively, to mix with the correlated fields $\hat a,~\hat b$, as shown in Fig. 3(b). Assume both BS1 and BS2 have the same transmissivity of $T$. The signal-transmitted outputs of the beam splitters:
\begin{eqnarray}
\hat c_{1} = \sqrt{T}\hat a_{01} + \sqrt{1-T}\hat a, ~~~~\hat c_{2} = \sqrt{T}\hat a_{02} - \sqrt{1-T}\hat b
\label{c12}
\end{eqnarray}
are combined with another 50:50 beam splitter. The added output of the 50:50 beam splitter is labeled as $c_{out}$:
$
\hat c_{out} = (\hat c_{1} + \hat c_{2})/\sqrt{2}.
$
Note that the signs in Eq. (\ref{c12}) are so chosen by adjusting relative phases between $\hat a_{01},~\hat a_{02}$ and between $\hat a,~\hat b$. Therefore, the output signal is simply $\langle X_{c_{out}}\rangle = 2\sqrt{T}(\langle X_{in}\rangle/\sqrt{2})/\sqrt{2} = 2\sqrt{T}\alpha$ and the noise is
\begin{eqnarray}
\Delta^2\hat X_{c_{out}} &=& \big\langle [\sqrt{T}(\Delta\hat X_{a_{01}} + \Delta\hat X_{a_{01}})/\sqrt{2} +\sqrt{1-T}(\Delta\hat X_{a} -\Delta\hat X_{b} )/\sqrt{2}]^2\big\rangle \cr
&=& T + (1-T)(\mu-\nu)^2.
\label{N1}
\end{eqnarray}
So the SNR for output $c$ is
\begin{eqnarray}
R_{c_{out}}^{(3)} = {T(2\alpha)^2 \over T + (1-T)(\mu-\nu)^2}.
\label{R31}
\end{eqnarray}
The superscript $(3)$ denotes the scheme for three-way splitting.

Now let us look at the other two outputs which are the two outputs of the amplifier:
\begin{eqnarray}
\hat a_{out} &=& G \hat a_1 + g\hat b_1^{\dag}\cr
\hat b_{out} &=& G \hat b_1 + g\hat a_1^{\dag}
\label{about}
\end{eqnarray}
with the two inputs $\hat a_1,~\hat b_1$ being the two signal-reflective outputs of BS1 and BS2, respectively:
\begin{eqnarray}
\hat a_{1} = -(\sqrt{T}\hat a -\sqrt{1-T}\hat a_{01}  ), ~~~~\hat b_{1} =  \sqrt{T}\hat b +\sqrt{1-T}\hat a_{02}.
\label{d12}
\end{eqnarray}
Here, an overall $\pi$-phase shift is added to $\hat a_1$ by adjusting its delay. So, the signals of outputs $\hat a_{out}$ and $\hat b_{out}$ are
\begin{eqnarray}
\langle \hat X_a \rangle &\equiv& \langle \hat a_{out} + \hat a_{out}^{\dag} \rangle = \alpha(G+g)\sqrt{2(1-T)} ,\cr
\langle \hat X_b \rangle &\equiv& \langle \hat b_{out} + \hat b_{out}^{\dag} \rangle = \alpha(G+g)\sqrt{2(1-T)} .
\label{X23}
\end{eqnarray}
The output noise can be calculated as
\begin{eqnarray}
\langle \Delta^2 \hat X_a \rangle &=& \langle (1-T)(G^2\Delta^2\hat X_{a_{01}} + g^2 \Delta^2\hat X_{a_{02}} ) + T (G\Delta\hat X_{a} - g\Delta\hat X_{b})^2\rangle ,\cr
\langle \Delta^2 \hat X_b \rangle &=& \langle (1-T)(g^2\Delta^2\hat X_{a_{01}} + G^2 \Delta^2\hat X_{a_{02}} ) + T (G\Delta\hat X_{b} - g\Delta\hat X_{a})^2\rangle .
\label{N23}
\end{eqnarray}
\begin{figure}[h]
\includegraphics[width=6in]{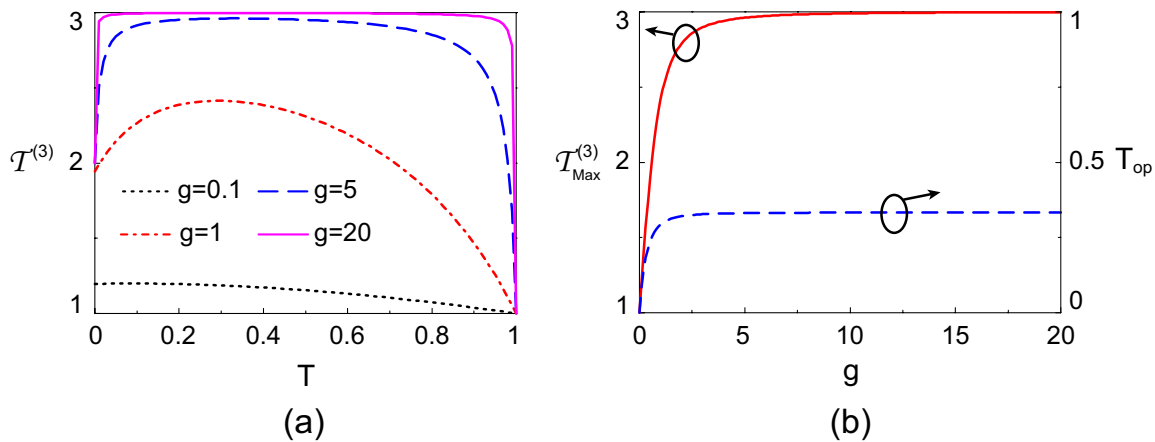}
\caption{(a) Overall transfer coefficient ${\cal T}^{(3)}$ for three-way quantum information splitting as a function of the transmissivity $T$ for different values of the gain parameter $g$ of NPA. (b) Optimized overall transfer coefficient ${\cal T}^{(3)}_{Max}$ and the optimum value of $T_{op}$ as a function of the gain parameter $g$.}
\end{figure}
It is straightforward to find
\begin{eqnarray}
\langle \Delta^2 \hat X_a \rangle = \langle \Delta^2 \hat X_b \rangle
 &=&(1-T)(G^2+g^2)
\cr &+&T(G^2+g^2)(\mu^2+\nu^2)-4TGg\mu\nu,
\label{N23-2}
\end{eqnarray}
which gives a minimum value
\begin{eqnarray}
\langle \Delta^2 \hat X_a \rangle = \langle \Delta^2 \hat X_b \rangle
 = (1-T)(G^2+g^2) +T,
\label{N23-m}
\end{eqnarray}
when $G=\mu, g=\nu$. So the SNRs for output $a$ and $b$ are
\begin{eqnarray}
R_a^{(3)}=R_b^{(3)} = {2 (G+g)^2 \alpha^2(1-T) \over (1-T)(G^2+g^2) +T}.
\label{R23}
\end{eqnarray}
With $R_{in} = (2\alpha)^2$ and $R_c^{(3)}$ from Eq.(\ref{R31}), we have the overall transfer coefficient for the three-way information splitting as
\begin{eqnarray}
{\cal T}^{(3)} &=& {R_a^{(3)} + R_b^{(3)}+ R_c^{(3)} \over R_{in}} \cr &=& {T (G+g)^2 \over T (G+g)^2 + (1-T)} + { (G+g)^2 (1-T) \over 2(1-T)g^2 +1 }.
\label{T3}
\end{eqnarray}
In the case when $g\gg1$, we have ${\cal T}^{(3)}\rightarrow 3$. This corresponds to ideal three-way splitting without noise added. For other values of $g$, we plot ${\cal T}^{(3)}$ in Fig. 4(a) as a function of the transmissivity $T$ for various values of $g$. For each value of $g$, when  $T=T_{op} \equiv g/(G+2g)$, ${\cal T}^{(3)}$ becomes the maximum value of \begin{eqnarray}
{\cal T}^{(3)}_{Max} =1+2g/G.
\label{Tmax}
\end{eqnarray}
In Fig. 4(b), we plot the maximum ${\cal T}^{(3)}_{Max}$ as well as the value of $T_{op}$ as a function of $g$. It can be seen that ${\cal T}^{(3)}_{Max} \rightarrow 3$ and $T_{op}$ is around 1/3 as $g \gg 1$.

In order to cascade the scheme for scaling up, we need $T\sim 1$. From Eq.(\ref{T3}), we find that as long as $2(1-T)g^2\gg 1$ or $g^2\gg {1\over2(1-T)}>1$, we have ${\cal T}^{(3)} \approx 3-{1\over(1-T)g^2} \rightarrow 3$. So, it is possible to achieve three-way noiseless splitting under the condition for cascade.

Notice that when $T=0$, i.e., there is no output at $\hat c_{out}$, this scheme is only a two-way information splitting and becomes the phase sensitive amplifier scheme for quantum information tapping \cite{gra}.

\section{Three-way quantum information Tap as a QND device}

Because of the involvement of the beam splitter and the amplification, the schemes discussed in previous sections cannot serve as an ideal QND measurement device. Holland et al discussed the criteria for a non-perfect QND measurement \cite{hol}. Let us now discuss how well our schemes can serve as a QND device along the line of argument of Ref. \cite{hol}.

We will not discuss the schemes in section 2 with only a beam splitter because Holland et al \cite{hol} already discussed the case with a squeezed state and the one with a correlated quantum source involves dual-beam and is not suitable for cascade required by a QND measurement. We will only concentrate on the scheme in Fig. 3(b) of section 4 with three-way splitting.

Since $\hat a_{out},~\hat b_{out}$ are amplified from the input $\hat a_{in}$, their individual noise is also much larger than vacuum noise so they are not likely to be the signal output of the non-demolition measurement but more likely to be the readout fields. So, $\hat c_{out}$ will be the signal output field. From Ref. \cite{hol}, the first criterion for QND measurement is the correlation coefficient between signal input and output field $C_{X_{a_{in}}X_{c_{out}}}$. It concerns how well the measurement scheme is for non-demolition. From the connection of $\hat c_{out}$ to the input $\hat a_{in}$, we find
\begin{eqnarray}
C_{X_{a_{in}}X_{c_{out}}} &=& {\langle \Delta X_{a_{in}}\Delta X_{c_{out}}\rangle\over \sqrt{\langle \Delta^2 X_{a_{in}}\rangle \langle \Delta^2 X_{c_{out}}\rangle }}\cr &=& {\sqrt{T}\over \sqrt{T +(1-T)(\mu-\nu)^2}} .
\label{Cin1}
\end{eqnarray}
This quantity is close to one for QND measurement if $(1-T)(\mu-\nu)^2 \ll T$ or $(\mu + \nu )^2 \gg (1-T)/T$, which can be realized with good correlation from the entangled fields. For cascade condition of $T\sim 1$, $(1-T)/T \sim 0$. So, it is easy to achieve this QND criterion even with modest quantum correlation (characterized by $\nu$). The second criterion is about how well the readout fields are correlated to the input field. This is quantified by the correlation coefficients:
\begin{eqnarray}
C_{X_{a_{in}}X_{a_{out}}} = C_{X_{a_{in}}X_{b_{out}}}  &=& {\langle \Delta X_{a_{in}}\Delta X_{a_{out}}\rangle\over \sqrt{\langle \Delta^2 X_{a_{in}}\rangle \langle \Delta^2 X_{a_{out}}\rangle }}\cr &=& {(G+g)\sqrt{(1-T)/2}\over \sqrt{(G^2+g^2)(1-T+T(\mu^2+\nu^2))-4TGg\mu\nu}}\cr &=& {(G+g)\sqrt{(1-T)/2}\over \sqrt{2(1-T)g^2 +1}} ~~~~{\rm for}~~G=\mu,~~g=\nu,
\label{Cin23}
\end{eqnarray}
which is approximately equal to $1-{T\over 4(1-T)g^2} \rightarrow 1$ if $2(1-T)g^2 \gg 1$ or $g^2 \gg {1\over 2(1-T)}$.

 Notice that
$|C_{X_{a_{in}}X_{a_{out}}}|^2 = R_a^{(3)}/ R_{in},~~~~|C_{X_{a_{in}}X_{b_{out}}}|^2 = R_b^{(3)}/ R_{in}, ~~~~|C_{X_{a_{in}}X_{c_{out}}}|^2 = R_c^{(3)}/ R_{in}$, so that they satisfy
\begin{eqnarray}
|C_{X_{a_{in}}X_{a_{out}}}|^2+|C_{X_{a_{in}}X_{b_{out}}}|^2+|C_{X_{a_{in}}X_{c_{out}}}|^2 = {\cal T}^{(3)} .
\label{C123T3}
\end{eqnarray}

The third criterion is about how well the scheme serves as a state projection device, which requires the conditional variance of output field on the readout fields to be smaller than one, the variance of the input coherent state. For Gaussian random variables, the conditional variance can be found as \cite{reid}
\begin{eqnarray}
V(X_{c_{out}}|X_{a_{out}}) &=& \langle (\Delta X_{c_{out}}-\lambda \Delta X_{a_{out}})^2\rangle_m\cr
&=& \langle \Delta^2X_{c_{out}}\rangle(1-C^2_{X_{c_{out}}X_{a_{out}}}).
\label{V12}
\end{eqnarray}
Here, subscript $m$ means that parameter $\lambda$ is optimized to minimize the value.
With
\begin{eqnarray}
&C_{X_{c_{out}}X_{a_{out}}} = {\langle \Delta X_{c_{out}}\Delta X_{a_{out}}\rangle\over \sqrt{\langle \Delta^2 X_{c_{out}}\rangle \langle \Delta^2 X_{a_{out}}\rangle }}\cr &= {(G+g)[1-(\mu-\nu)^2]\sqrt{T(1-T)/2}\over \sqrt{[T +(1-T)(\mu-\nu)^2][(G^2+g^2)(1-T+T(\mu^2+\nu^2))-4TGg\mu\nu]}}\cr &= {(G+g)[1-(G-g)^2]\sqrt{T(1-T)/2}\over \sqrt{[T+(1-T)(G-g)^2][2g^2(1-T)+1]}} ~~{\rm for}~~G=\mu,~~g=\nu ,
\label{C12}
\end{eqnarray}
and
$\langle \Delta^2 X_{c_{out}}\rangle =T + (1-T)(G-g)^2$ when $G=\mu, g=\nu$,
the final result is
\begin{eqnarray}
V(X_{c_{out}}|X_{a_{out}}) &=& {1-T\over (G+g)^2} + {T\over 2(1-T)g^2 +1 } \cr
&\approx & {1+T^2\over 4g^2(1-T)} ~~~~{\rm for}~~g^2\gg {1\over 2(1-T)}.
\label{V12-1}
\end{eqnarray}
$V(X_{c_{out}}|X_{b_{out}})$ is the same as the above. The conditions for these two quantities smaller than one and close to zero are the same as the conditions for the first two QND criteria discussed earlier. So, this scheme can serve as a QND device when $g^2\gg {1\over 2(1-T)}$.

Since there are two readouts, they should also be well correlated \cite{ben95}, which can be quantified by the correlation coefficient:
\begin{eqnarray}
C_{X_{a_{out}}X_{b_{out}}} &=& {\langle \Delta X_{a_{out}}\Delta X_{b_{out}}\rangle\over \sqrt{\langle \Delta^2 X_{a_{out}}\rangle \langle \Delta^2 X_{b_{out}}\rangle }}\cr &=& {2Gg(1-T)\over 2(1-T)g^2 +1} ~~~~{\rm for}~~G=\mu,~~g=\nu ,
\label{Cab}
\end{eqnarray}
which is approximately equal to $1-{T\over 2(1-T)g^2} \rightarrow 1$ for $g^2 \gg {1 \over 2(1-T)}$.

The state projection is quite unique for quantum measurement. For non-ideal case, this is reflected as the third in the criteria by Holland et al \cite{hol}. In the ideal quantum measurement, because the state is projected to the eigen-state of the measurement, subsequent measurement will give out the same result as the first one. For non-ideal case, however, the projected state is not exactly the eigen-state of the measurement. For a good QND measurement scheme, subsequent measurement should project to a state that is closer to the eigen-state than the first one.
This property is not discussed by Holland et al. So, for sequential non-ideal QND measurement, a fourth criterion should be required that addresses the additional information with two readouts. This was discussed in Ref. \cite{bru97}. Here we rephrase it in terms of conditional variance. Since the conditional variance $V(X_{c_{out}}|X_{a_{out}})$ on one measurement outcome is used to characterize the state projection quality of one non-ideal QND measurement, we should use the conditional variance $V(X_{c_{out}}|X_{a_{out}},X_{b_{out}})$ on two measurement outcomes for sequential non-ideal QND measurement. A good state projection should require $V(X_{c_{out}}|X_{a_{out}},X_{b_{out}}) < V(X_{c_{out}}|X_{a_{out}})$. We can find $V(X_{c_{out}}|X_{a_{out}},X_{b_{out}}) $ in a similar way as $ V(X_{c_{out}}|X_{a_{out}})$ by minimizing the quantity: $\langle (\Delta X_{c_{out}}-\lambda_a \Delta X_{a_{out}}-\lambda_b \Delta X_{b_{out}})^2\rangle_m$. The result is
\begin{eqnarray}
V(X_{c_{out}}|X_{a_{out}},X_{b_{out}}) &=& {1-T\over (G+g)^2} + {T+2T(1-T)g/(G+g)\over 2(1-T)g(G+g) +1 }  \cr &\approx & {1\over 4g^2(1-T)} ~~~~{\rm for}~~g^2 \gg {1\over 4(1-T)}.
\label{V123}
\end{eqnarray}
Comparing Eq. (\ref{V123}) with Eq. (\ref{V12-1}), we find for $g^2 \gg {1\over 2(1-T)} > {1\over 4(1-T)}$
\begin{eqnarray}
{V(X_{c_{out}}|X_{a_{out}},X_{b_{out}}) \over V(X_{c_{out}}|X_{a_{out}})} \approx {1\over 1+T^2} < 1.
\label{V123-2}
\end{eqnarray}
Thus, the criterion for sequential QND measurement is satisfied.

\section{Summary and discussion}

In summary, we proposed a scheme of three-way quantum information tapping by combining a beam splitter scheme with a quantum correlated noiseless amplification scheme, in which the two readout fields are amplified while the third output is almost unchanged that can be cascaded down for further information distribution. At large gain of the amplification, the overall information transfer coefficient is close to the optimum quantum value of 3, corresponding to noiseless information splitting. Such a three-way information tapping scheme, satisfying a set of four criteria for QND measurement,  can  serve for a sequential QND measurement.

Compared to the beam splitter scheme with a squeezed state or a correlated source, at the same level of signal attenuation (by a factor of the beam splitter transmissivity T), we achieved two readout outputs instead of one. Thus, our scheme is more efficient in quantum information tapping.  In all the cases discussed, ideal conditions are reached at high gain level which is quantified as $2g^2(1-T) \gg 1$. This requirement corresponds to the two amplified readout signals are larger than the original input signal. This is the advantage of the current scheme involving noiseless amplification: it can give rise to appreciable readout signal levels even if the portions initially distributed to these readout channels  are small ($1-T \ll 1$ if $T\sim 1$).

In our scheme of three-way quantum information tapping presented in Fig. 3(b), there is a vacuum input $\hat a_v$ at the 50:50 beam splitter for input signal splitting.
What role does it play? Actually, it does not contribute to  $\hat c_{out}$ field at all: with the proper phase, the contribution of $\hat a_v$ is canceled at $\hat c_{out}$ field so it will come out at the other output port of the 50:50 combining beam splitter (which will not contain any information about the input, either, by the same argument).
On the other hand, the vacuum input $\hat a_v$ does contribute to the two amplified readouts: $\hat a_{out},~\hat b_{out}$.
After a detailed look, however, we find that the contribution of $\hat a_v$ is deamplified and becomes negligibly small for a large $g$.

This work was supported in part by the National NSF of China (No. 11527808), the State Key Development Program for Basic Research of China (No. 2014CB340103), the Specialized Research Fund for the Doctoral Program of Higher Education of China (No. 20120032110055), The Natural Science Foundation of Tianjin (No. 14JCQNJC02300), PCSIRT and 111 Project B07014.

\end{document}